\documentclass[12pt,aps,nofootinbib,pra]{revtex4}   
\usepackage{amsmath}    
\usepackage{amsfonts}
\usepackage{graphicx}   
\usepackage{subfigure}
\begin{document}

\title{Visualizing Imaginary Rotations and Applications in Physics}
\author{Eli Lansey}
\affiliation{Department of Physics, City College and The Graduate Center of the City University of New York, New York, NY 10031}
\email{elansey@gc.cuny.edu}
\date{June 5, 2009}

\begin{abstract}
I discuss the notions of traditional vector length, and suggest defining a complex vector length for complex vectors, as opposed to the traditional Hermitian real length.  The advantages of this are shown in the development of rotations through imaginary angles.  Emphasis is placed on visualizing these quantities and rotations graphically, and I show some applications in physics: Lorentz transformations, Grassmann variables, and Pauli spin matrices.

\end{abstract}

\maketitle

\newcommand{\ip}[1]{\left(#1,#1\right)}

\section{Introduction}
In 1962, in the first edition of his seminal \textsl{Classical Electrodynamics}, J.D. Jackson suggests viewing Lorentz transformations ``as orthogonal transformations in four dimensions,''~\cite{jackson1} where the time coordinate is chosen to be an imaginary quantity and the three spacial coordinates are real.  He shows that one can consider a Lorentz transformation as a rotation of axes through an imaginary angle $\psi$, and tries to show this graphically with traditional rotation-of-axes diagrams.  However, as he notes, these graphs are not the most ideal ways to show Lorentz transformations, as $\cos\psi\geq 1$.  He therefore concludes that ``the graphical representation of a Lorentz transformation as a rotation is merely a formal device,'' and leaves it at that.  
In fact, in the two later editions of his text this discussion is removed completely.




In this paper I revisit Jackson's original idea, but show that we need to carefully rethink the notions of length, angles and rotations. I show that, if we draw the right pictures, it is straightforward to visualize arbitrary complex vectors, not just those of the form described by Jackson.   Additionally, I show how to visualize rotations of these vectors through real, as well as imaginary angles.  I have also developed a set of computer tools using Mathematica to help with these visualizations, available online.

The discussion also highlights the graphical distinction between unitary and orthogonal matrices.  This distinction is highlighted in the physical applications; Pauli spin matrices behave differently than rotation matrices.  Furthermore, I develop a way of visualizing non-commuting Grassmann numbers using this broader methodology.




\section{Vectors and Length}
Before we can begin thinking about rotations and their mathematical and graphical properties, we first need to discuss vectors, both real and complex, and the notion of vector length.
\subsection{Real Vectors}

\begin{figure}
	\centering
		\includegraphics[width=0.50\textwidth]{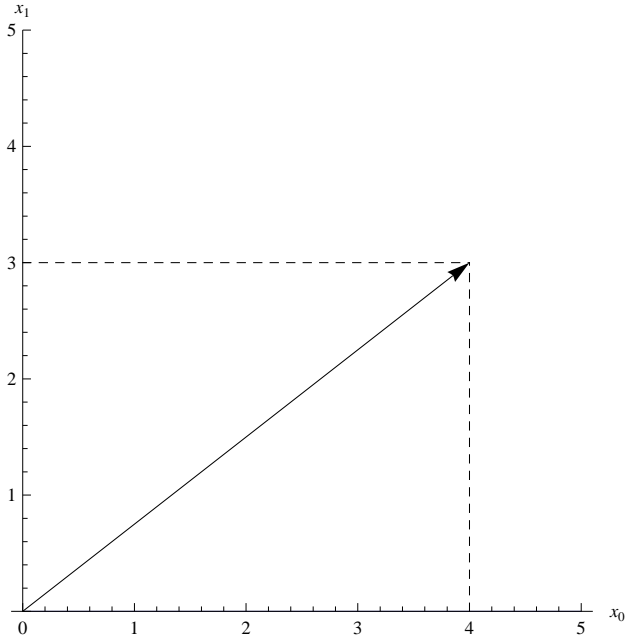}
	\caption{Visualizing a real 2D vector $\vec{x}=(4,3)$ in the $x_0$$x_1$-plane.  $r=5$ for this example.}
	\label{fig:realvec}
\end{figure}

Consider a traditional 2D vector
\begin{equation}
\label{rvec}
\vec{x}=\begin{pmatrix}
x_0 \\
x_1  \hfill
\end{pmatrix},
\end{equation}
where $x_0$ and $x_1$ are real numbers.\footnote{I chose to index these components starting from 0, rather than 1 as is traditional for considerations which come in Section~\ref{sec:relativity}}  We can easily visualize this on a 2D graph, see Fig.~\ref{fig:realvec}.  In this case we can ``see'' the length $r$ of this vector, and use the standard inner product
\begin{equation}
\label{innerprod}
\ip{\vec{x}}=\vec{x}^{\,\mathrm{T}}\vec{x}
\end{equation}
to find its value:
\begin{equation}
r^2=\ip{\vec{x}},
\end{equation}
or
\begin{equation}
\label{rinnerprod}
r^2 =  x_0^2+x_1^2,
\end{equation}
for our 2D example.\cite{lay}  Since $x_0$ and $x_1$ are real values, both $x_0^2$ and $x_1^2$ are larger than zero, giving Cauchy's inequality
\begin{equation}
\label{cauchy}
\ip{\vec{x}}\geq 0.
\end{equation}

Thus, we'll \emph{define} the length $r$ of an $n$-dimensional vector $\vec{x}$ as:
\begin{equation}
\label{rlength}
r\equiv\sqrt{\ip{\vec{x}}}
\end{equation}
or
\begin{equation}
r=\sqrt{x_0^2+x_1^2}
\end{equation}
in our 2D case, where we've chosen the positive root as a matter of convention.

Then, this length has all the physical properties of length that we are used to:  It is greater than, or equal to zero, it is what you get by physically measuring with a ($n$-dimensional) ruler, and so on.  So I will forgo further discussion of these familiar real lengths and vectors for the moment and proceed on to complex vectors.

\newcommand{\vx}{{\boldsymbol{\vec{x}}}}
\newcommand{\x}[2]{x_{#1}^{#2}}
\newcommand{\xx}[1]{x_{#1}^{r}+i\,x_{#1}^i}

\subsection{Complex Vectors}
\begin{figure}
    \centering
    \subfigure[ Real component of $\vx$, $\vec{x}\,^r$]{\label{fig:rvx}\includegraphics[width=0.3\textwidth]{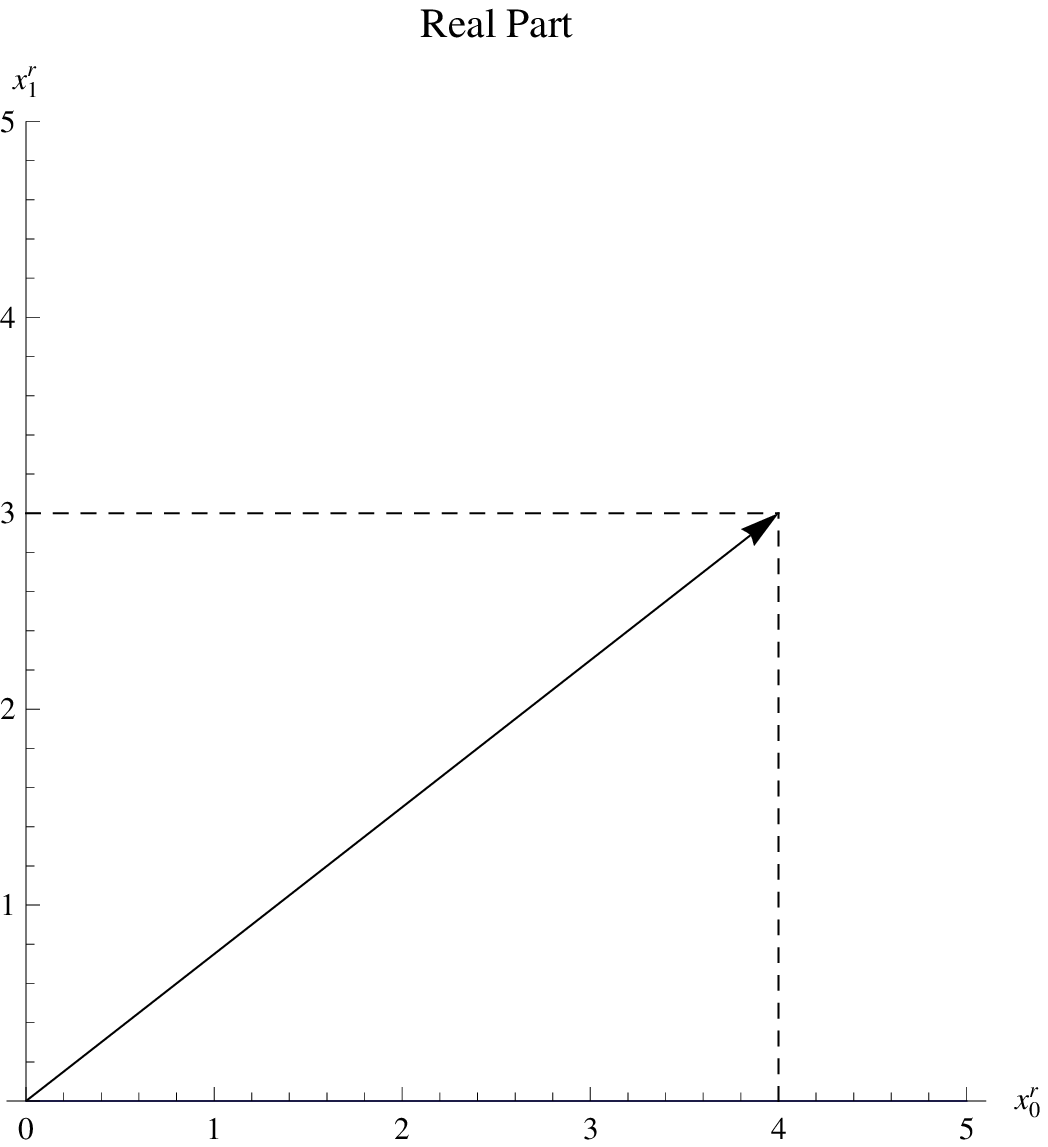}}
    \subfigure[ Imaginary component of $\vx$, $\vec{x}\,^i$]{\label{fig:cvx}\includegraphics[width=0.3\textwidth]{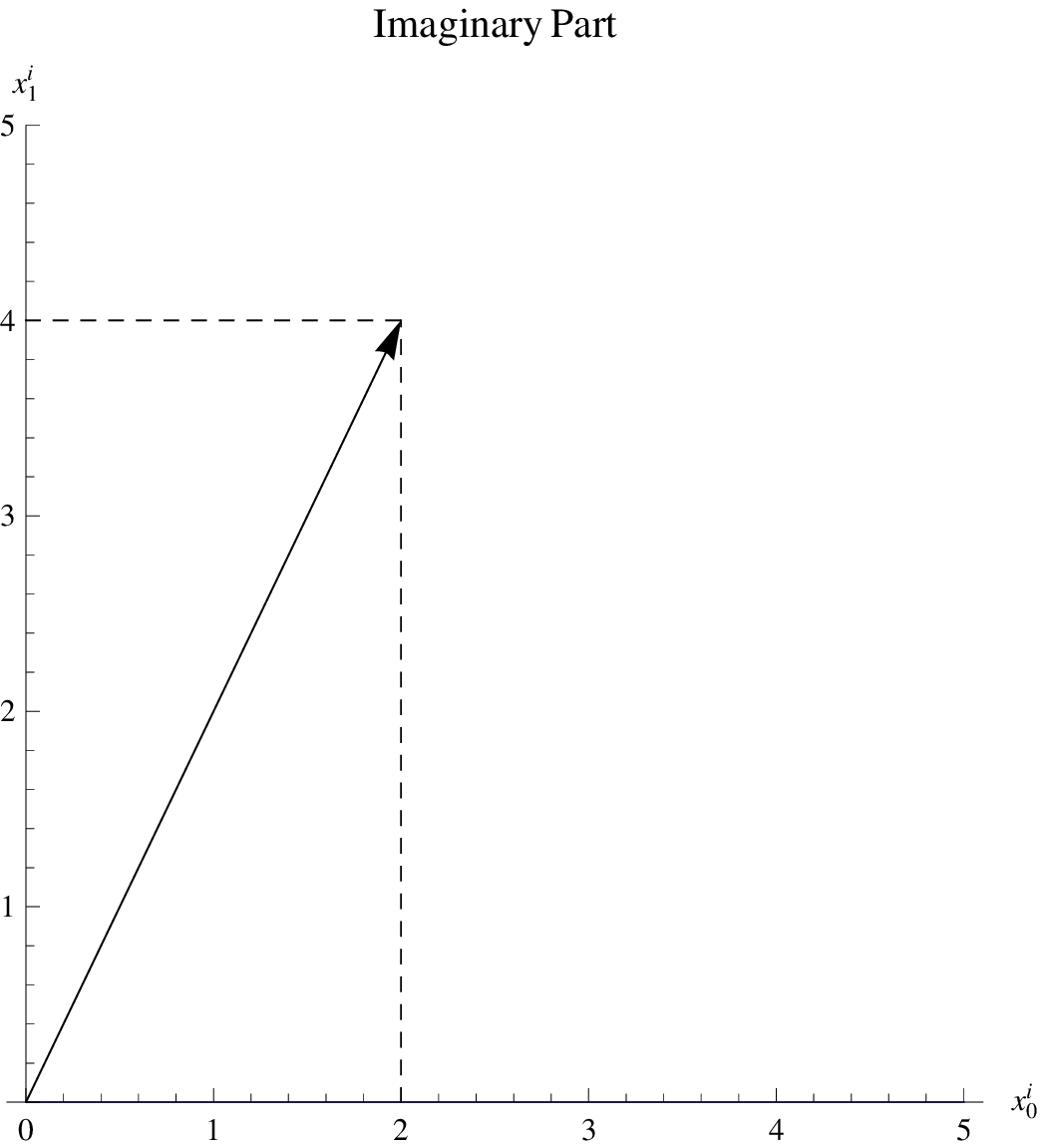}}
    \caption{Visualizing a complex 2D vector $\vx=(4+2i,3+4i)$ by looking separately at $\vec{x}\,^r=(4,3)$ in the $x_0^r$$x_1^r$-plane in~\subref{fig:rvx} and $\vec{x}\,^i=(2,4)$ in the $x_0^i$$x_1^i$-plane in~\subref{fig:cvx}.  $h=3\sqrt{5}$ and $s=\sqrt{5}$ for this example.}
	\label{fig:cvec}
\end{figure}

We'll define a complex 2D vector $\vx$ (the boldface signifies a complex vector) as a vector whose components are now allowed to be complex numbers:
\begin{equation}
\label{cvec}
\vx=\begin{pmatrix}
\xx{0} \\
\xx{1} \hfill
\end{pmatrix},
\end{equation}
where $x_0^r$, $x_0^i$, $x_1^r$ and $x_1^i$ are all real numbers.  To help visualize this quantity, we note that we can rewrite this as
\begin{equation}
\label{vis}
\vx=\begin{pmatrix}
x_0^r  \\
x_1^r \hfill
\end{pmatrix}+i\, \begin{pmatrix}
x_0^i \\
x_1^i \hfill
\end{pmatrix}=\vec{x}\,^r+i\,\vec{x}\,^i,
\end{equation}
where $\vec{x}\,^r$ and $\vec{x}\,^i$ are both real vectors.\cite{gibbs2}  This allows us to visualize the vector as two different vectors, one corresponding to the real part of the vector, the other corresponding to the complex part. See Fig.~\ref{fig:cvec}.

\label{polar}This is, of course, not the only way of visualizing complex vectors.  We could also plot different Argand diagrams, one for each of the $x_i$ components.  This second approach is beneficial in the sense that it's easier to plot higher dimensions, however is disadvantageous in that these plots do not capture the spacial orientation of these vectors.  Additionally, for other cases, a polar visualization -- plotting the magnitudes $|x_0|$ vs $|x_1|$ on one plot, and the phases on the other -- is more helpful.  This is especially useful when we are really only interested in either the real or imaginary parts of the vector, but keep complex notation for convenience, such as with complex electric and magnetic fields.  In this case, however, I believe the first approach is the most natural way to visualize these vectors.

The trouble now is in defining a length.  On one hand, we'd like to keep Cauchy's inequality (Eq.~\ref{cauchy}) and consider a real length of complex vector.  A traditional way toward this end is through the so-called Hermitian inner product:
\begin{equation}
\left<\vx,\vx\right>\equiv\vx^{\,\dagger}\,\vx
\end{equation}
where $\vx^{\,\dagger}\equiv \left(\vx^{\,*}\right)^{\,\mathrm{T}}$, i.e. the complex transpose of $\vx$ such that
\begin{equation}
h^2=\left<\vx,\vx\right>
\end{equation}
or
\begin{eqnarray}
h^2&=&\left(\xx{0}\right)^*\left(\xx{0}\right)+\left(\xx{1}\right)^*\left(\xx{1}\right)\\
&=&\left(x_0^r\right)^2+\left(x_0^i\right)^2+\left(x_1^r\right)^2+\left(x_1^i\right)^2
\end{eqnarray}
in our 2D case.  This, indeed, gives a number $h^2>0$, corresponding the sum of the squares of the (real) lengths of $\vec{x}\,^r$ and $\vec{x}\,^i$, satisfying Cauchy's inequality~(Eq.~\ref{cauchy}), and allowing us to \emph{define} a real length $h$ of an imaginary vector as we did in Eq.~\ref{rlength}:
\begin{equation}
\label{hlength}
h\equiv\sqrt{\left<\vx,\vx\right>},
\end{equation}
or
\begin{equation}
h=\sqrt{\left(x_0^r\right)^2+\left(x_0^i\right)^2+\left(x_1^r\right)^2+\left(x_1^i\right)^2}
\end{equation}
in our 2D case.\cite{larson}

This length has its advantages.  First and foremost, it is a \emph{real} length, with all of real length's properties.  It can, in principle, be measured, it is greater than or equal to zero, and so on.  A variation on this length is often used, for example, in quantum mechanics to find real expectation values of measurable physical quantities from a complex wavefunction.  Thus, the major benefit of the Hermitian inner product is that it allows us to extract a real length from a complex vector.

The trouble is, we lose all information about the complex nature of the vector.  It is as if the $i$ doesn't even exist.  Consider, for example, the purely imaginary vector $\vx=(i,0)$.  The Hermitian inner product gives $h^2=1$, or $h=1$, the same as the product from a purely real vector $\vx=(1,0)$.  But, in some sense, the length of the vector is, in fact, imaginary, in that it exists purely in the imaginary part of the complex 2D domain, contrasted with the real length of a purely real vector which exists purely in the real 2D domain.

Therefore, we'll try to develop an alternative length which carries with it the complexity of the vector space.  To start, we'll study what happens to the traditional inner product if we feed it a complex vector $\vx$ instead of a real vector.  Again, consider, for example, the purely complex vector $\vx=(i,0)$.  Putting it through the machinery of the inner product (Eq.~\ref{innerprod}) we get
\begin{equation}
\ip{\vx}=\left(0+i\right)^2+\left(0+i\,0\right)^2=-1
\end{equation}
This inner product clearly violates Cauchy's inequality~(Eq.~\ref{cauchy}), which, for some reason, sends some people into conniptions.
 But let's continue anyway, in spite of this reservation, and \emph{define}, as we did in Eq.~\ref{rlength}, a complex ``length'' $s$:
\begin{equation}
\label{clength}
s\equiv\sqrt{\ip{\vx}},
\end{equation}
or
\begin{equation}
s=\sqrt{\left(\xx{0}\right)^2+\left(\xx{1}\right)^2}
\end{equation}
in our 2D case, where, again, we've chosen the positive root as a matter of convention.  But since we can potentially have $\ip{\vx}<0$, $s$ can be a complex number!\footnote{If the idea of a complex length seems strange to you, consider the opposition to complex numbers when people first suggested their existence.}  As an example, consider, once again, our old friend, the purely imaginary vector $\vx=(i,0)$.  Its complex length is $s=i$.  This now carries the information that the vector is actually complex, and can be contrasted with $\vx=(1,0)$ whose length is $s=1$.

The notion of a complex length is not completely unheard of in mathematics literature.  See for example, Dodson and Poston,\cite{tensorgeom} where they entertain the option of such a length in the Minkowski metric.  They ultimately reject this length, regarding the inner product itself as more important, since, within the framework of the Minkowski metric, all quantities are real.  In this case however, the quantities are essentially complex, and, as such, a complex length is no longer adventitious.

\label{seq0}The behavior of this length $s$ is a bit nasty, though.  Aside from the obvious problem of measuring with a complex ruler,\footnote{Not that this is inherently any stranger than measuring with an $n$-dimensional ruler.} $s$ can be entirely real or entirely complex for vectors that are part real and part complex, and can equal zero for non-zero vectors.  For example, for $\vx=(5i,3)$, $s=4i$, but for $\vx=(3i,5)$, $s=4$; for $\vx=(1+i,-1+i)$, $s=0$.  I discuss the case of $s=0$ later, in Section~\ref{sec:grassman}.  Additionally, the imaginary part of $s$ can be less than 0.  
In general, actually, the length of $\vx^*$ is the conjugate of the length of $\vx$.

In any case, ultimately, the length $s$ of $\vx$ is a single complex number, which can, of course, be visualized according to your favorite method.  So then, we have three different lengths:
\begin{itemize}
\item $r=\sqrt{\ip{\vec{x}}}$, real length for real vectors
\item $h=\sqrt{\left<\vx,\vx\right>}$, real length for complex vectors
\item $s=\sqrt{\ip{\vx}}$, complex length for complex vectors
\end{itemize}
And, although $s$ is tricky to get to know, its significance, and the importance of the difference between $s$, $h$ and $r$ will be highlighted in Section~\ref{sec:complexrot}.

\section{Rotations}
\subsection{Real Rotations}
\begin{figure}[]
	\centering
		\includegraphics[width=0.50\textwidth]{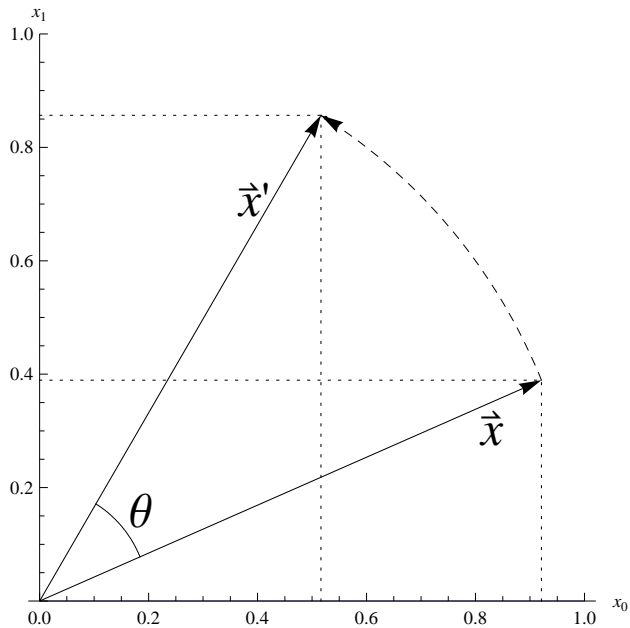}
	\caption{Traditional visualization of the rotation of a (unit) vector $\vec{x}$ through angle $\theta$ to $\vec{x}'$.}
	\label{fig:rrotr}
\end{figure}

Say we want to rotate the real, 2D vector vector $\vec{x}$ through some angle $\theta$ around a vector perpendicular to the 2D plane.  See Fig.~\ref{fig:rrotr}.
The resulting vector
\begin{equation}
\vec{x}'=\begin{pmatrix}
x_0' \\
x_1'  \hfill
\end{pmatrix},
\end{equation}
has components which are linear combinations of the original components $x_0$ and $x_1$.  An important, nearly defining characteristic of any such a rotation is that is preserves the length $r$ of the vector, given by equation~(Eq.~\ref{rlength}).  

Using Fig.~\ref{fig:rrotr} we can write $x_0'$ and $x_1'$ in terms of $\theta$ and the initial components $x_0$ and $x_1$ and construct the standard 2D matrix of rotation:
\begin{equation}
\label{matrix}
R(\theta)=\begin{pmatrix}
\cos \theta \hfill & -\sin \theta \\
\sin \theta \hfill & \cos \theta
\end{pmatrix},
\end{equation}
and write
\begin{equation}
\label{rmap}
\vec{x}'=R(\theta)\vec{x}
\end{equation}
Easy calculations verify that $r=r'$.  Additionally, since this is a real vector, $r=h=s=r'=h'=s'$, so a distinction between the various lengths does not arise.

\begin{figure}
    \centering
    \includegraphics[width=0.9\textwidth]{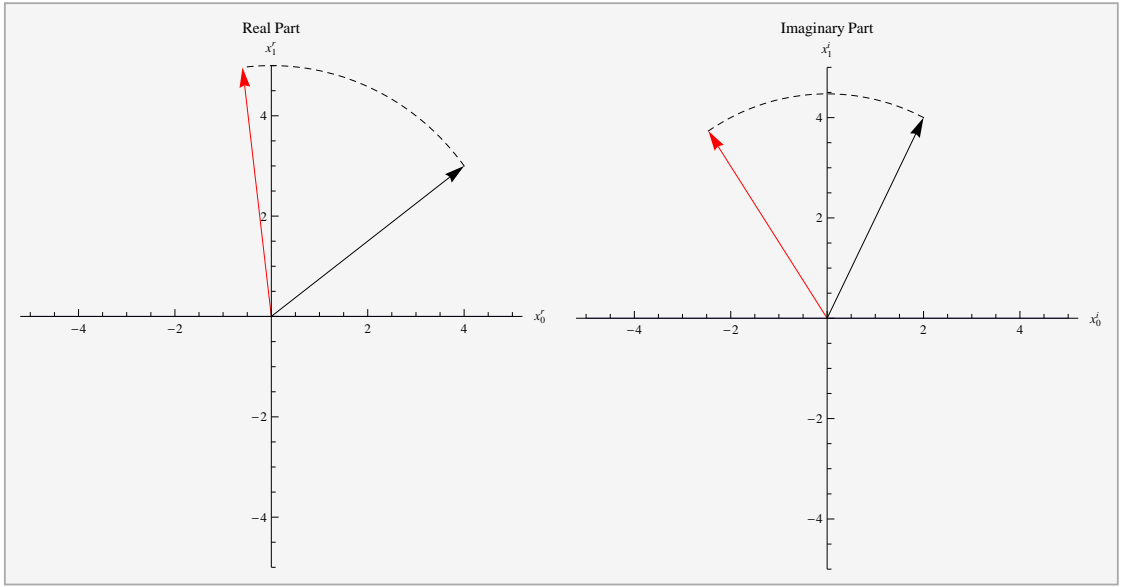}
    \caption{Visualizing real rotation of a complex 2D vector $\vx=(4+2i,3+4i)$ through angle $\theta=\pi/3$ by looking separately at the rotations of $\vec{x}\,^r=(4,3)$ in the $x_0^r$$x_1^r$-plane and $\vec{x}\,^i=(2,4)$ in the $x_0^i$$x_1^i$-plane. Black arrow is original vector, red arrow is rotated vector, dotted line is path traced under rotation}
	\label{fig:rrotc}
\end{figure}

\label{sec:rrotcv}
Now, suppose we want to rotate the complex vector vector $\vx$ through some angle $\theta$ around a vector perpendicular to the complex 2D plane.  Nothing is stopping us from using Eq.~\ref{rmap} with a vector $\vx$, so we may as well consider $\vx'=R(\theta)\vx$.  We will use Eq.~\ref{vis} to help us visualize this rotation:
\begin{subequations}
\begin{eqnarray}
\vx' &=&
\label{eq:xmapr}
  R(\theta)\vec{x}\,^r+i\,R(\theta)\vec{x}\,^i \\
\label{eq:hmapr}
  &=&Q^{(r)} \begin{pmatrix}
\cos\theta  \hfill \\
\sin\theta  \hfill
\end{pmatrix} \hfill + i\,Q^{(i)}\begin{pmatrix}
\cos\theta  \hfill \\
\sin\theta  \hfill
\end{pmatrix}
\end{eqnarray}
\end{subequations}
where
\begin{subequations}
\label{eq:qs}
\begin{eqnarray}
Q^{(r)} &\equiv& \begin{pmatrix}
x_0^r \hfill & -x_1^r \\
x_1^r \hfill & x_0^r
\end{pmatrix}\\
Q^{(i)} &\equiv& \begin{pmatrix}
x_0^i \hfill & -x_1^i \\
x_1^i \hfill & x_0^i
\end{pmatrix}.
\end{eqnarray}
\end{subequations}
In other words, there are two equally valid graphical ways of thinking about and visualizing this rotation.  The first~(Eq.~\ref{eq:xmapr}) views it as an identical rotation of the real and imaginary components of $\vx$ through an angle $\theta$ whereas the second~(Eq.~\ref{eq:hmapr}) views it as a mapping of a portion of the unit circle (up to angle $\theta$) due to the real and imaginary components of $\vx$.  The second one emphasizes that the path traced by the vector under this rotation is a circle.
This is also seen clearly from Eq.~\ref{rinnerprod}, which is the equation of a circle of radius $r$.  Graphically, we can see the equivalence of both viewpoints, see Fig.~\ref{fig:rrotc}.

Since this is a rotation, we need to check that length is conserved through this operation.  In this case, carrying through calculations for an arbitrary $\vx$ and $\theta$, even though $s\neq h$, $h=h'$ and $s=s'$.  In other words, both the Hermitian length and the complex length of $\vx$ are conserved through this rotation.  Here again, then, an important distinction between these two lengths does not arise.

\subsection{Imaginary Rotations}
\label{sec:complexrot}
Having understood, and visualized real rotations of complex vectors, let's see what happens if we rotate $\vx$ through an imaginary angle $i\theta$.  To do this, we'll make use of Eq.~\ref{rmap}, and just substitute the angle $i\theta$ for $\theta$.  So, first, let's see what happens to $R(\theta)$.  Making use of Eq.~\ref{matrix} we write:
\begin{equation}
\label{cmatrix}
R(i\theta)=\begin{pmatrix}
\cos i\theta \hfill & -\sin i\theta \\
\sin i\theta \hfill & \cos i\theta
\end{pmatrix}=
\begin{pmatrix}
\cosh \theta \hfill & -i \sinh \theta \\
i \sinh \theta \hfill & \cosh \theta
\end{pmatrix}
,
\end{equation}
where we use the standard transformation to hyperbolic trigonometric functions.\cite{needham}  Since I claim this is a rotation, we need to verify that length is conserved.  In this case, carrying through calculations for an arbitrary $\vx$ and $i\theta$ shows that $s=s'$ (see Appendix~\ref{dix:cons}), whereas $h\neq h'$.  Thus, it is the complex length of a vector that is conserved under any imaginary rotation.  Additionally, note that while real rotations are periodic in $2\pi$, imaginary rotations are not; one can keep imaginarily rotating without ever returning to one's initial orientation.

Looking at the form of $R(i\theta)$ given by Eq.~\ref{cmatrix}, I would like to make the following strange statement:
\begin{quote}
\label{qt:hyprot}
Rotation through an imaginary angle $i\theta$ can be understood as a \emph{hyperbolic rotation} through a real angle $\theta$.
\end{quote}
But what do I mean by a hyperbolic rotation and how might we visualize it?  And what is the geometrical meaning of the angle $\theta$ in this case?

\subsubsection{Ordinary Rotation}
First, we need to rethink regular rotations.  What do we mean by rotation through an angle $\theta$?  What exactly is $\theta$ and how can we visualize it?

\begin{figure}
    \centering
		\includegraphics[width=0.50\textwidth]{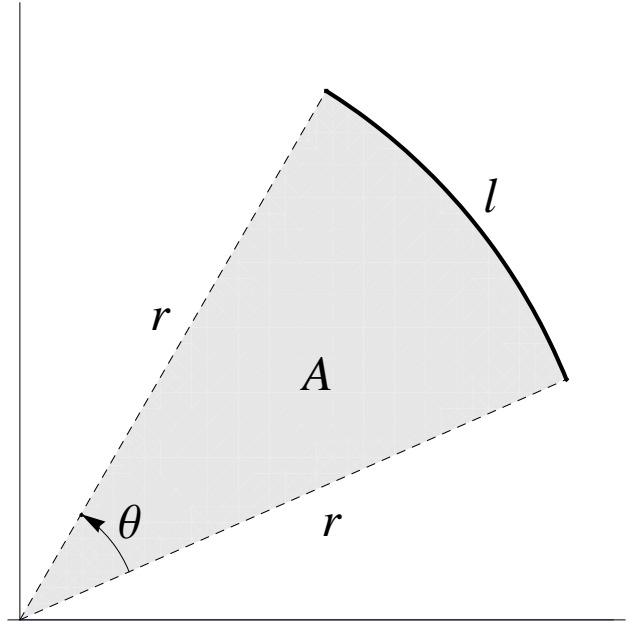}
    \caption{Relationship between arclength $l$, radius $r$, area $A$ and $\theta$}
	\label{fig:arclength}
\end{figure}

Typically, we define the unitless angle $\theta$
\begin{equation}
\label{eq:arclength}
\theta\equiv\frac{l}{r}
\end{equation}
where $l$ is the arclength along an arc of radius $r$, as in Fig.~\ref{fig:arclength}.  To find the angle of a full circle we simply substitute the empirical expression $C=2\pi r$ for the circumference, $C$, of a circle of radius $r$ in place of the arclength $l$ in Eq.~\ref{eq:arclength}:
\begin{equation}
\theta_{full circle}=\frac{2\pi r}{r}=2\pi
\end{equation}
Then, say we wanted to know the measure of the angle that sweeps $\frac{1}{4}$ of a circle.  To find this angle, we'd note that $l_{quarter circle}=\frac{1}{4}l_{full circle}=\frac{1}{4}C$ which can be substituted back into Eq.~\ref{eq:arclength}.  Thus,
\begin{equation}
\theta_{quarter circle}=\frac{1}{4}2\pi=\frac{\pi}{2}.
\end{equation}
In other words, Eq.~\ref{eq:arclength} defines the angle in terms of \emph{unitless fractions of the circumference} of a circle.  But, in principle, any method which provides an alternate means of measuring a \emph{unitless fraction of a circle} can be used as the basis of an angle definition.  For example, the commonly used degree measure is simply a count of $\frac{1}{360}$ths of a circle.

\begin{figure}
    \centering
		\includegraphics[width=0.50\textwidth]{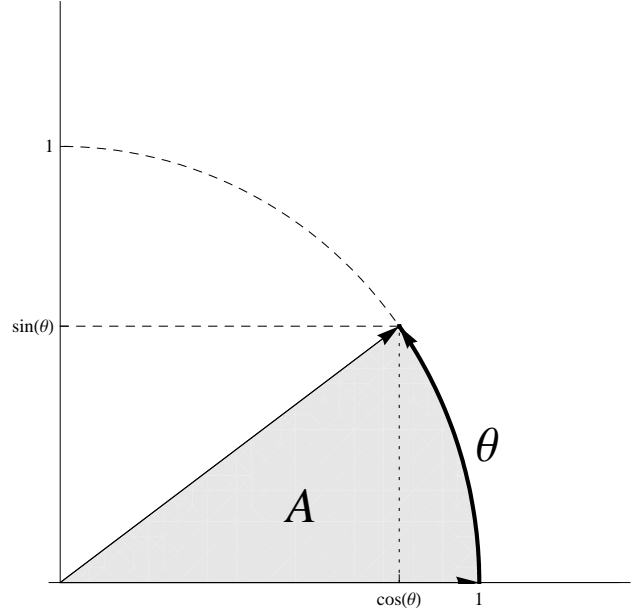}
    \caption{Relationship between $\theta$ and area $A$ in a unit circle}
	\label{fig:area}
\end{figure}

So, we'll provide an alternate definition in terms of fractional \emph{area} of a circle:
\begin{equation}
\label{area}
\theta\equiv\frac{2A}{r^2}
\end{equation}
where $A$ is the shaded area shown in Fig.~\ref{fig:arclength}.\footnote{A natural extension to 3D expresses solid angle $\Omega$ in terms of fractional (solid) volume, $V$: $\Omega=3V/r^3$}  Then, to find the angle of a full circle, we substitute the emprical expression $A=\pi r^2$ for the area of a circle of radius $r$ into (Eq.~\ref{area}) and get $\theta_{full circle}=2\pi$, and so on.  Since we are, ultimately, only interested in fractions of a circle, and to make life simpler ahead, we'll adjust the definition given in Eq.~\ref{area} slightly, and define $\theta$ in terms of fractional area of the unit circle of radius $r=1$, given by
\begin{equation}
x^2+y^2=1
\end{equation}
or, alternately, by the parametric equation,
\begin{subequations}
\label{rrotparam}
\begin{eqnarray}
x(\theta)&=&\cos\theta \\
y(\theta)&=&\sin\theta
\end{eqnarray}
\end{subequations}
Then Eq.~\ref{area} becomes
\begin{equation}
\label{carea}
\theta=2A
\end{equation}
where $A$ is the area shown in Fig.~\ref{fig:area}.

\subsubsection{Hyperbolic Pseudo-Rotation}
\begin{figure}
    \centering
		\includegraphics[width=0.50\textwidth]{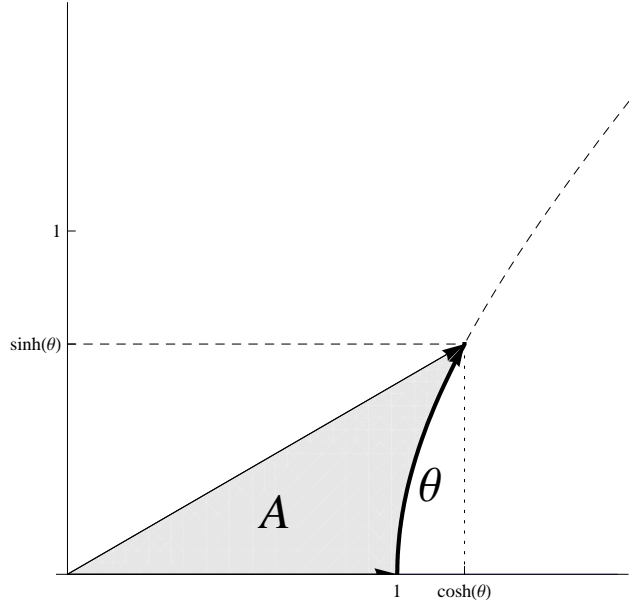}
    \caption{Relationship between $\theta$ and area $A$ in a unit hyperbola}
	\label{fig:harea}
\end{figure}
Now we can try hyperbolic rotation.  Here we'll consider the unit hyperbola given by
\begin{equation}
x^2-y^2=1
\end{equation}
or, alternately, by the parametric equation
\begin{subequations}
\label{hyprotparam}
\begin{eqnarray}
x(\theta)&=&\cosh\theta \\
y(\theta)&=&\sinh\theta.
\end{eqnarray}
\end{subequations}
Then, just as with the circle (Eq.~\ref{carea}),
\begin{equation}
\label{harea}
\theta=2A
\end{equation}
where $A$ is the area shown in Fig.~\ref{fig:harea}.  Note the similarity between Eqs.~\ref{hyprotparam}~and~\ref{rrotparam}.  This motivates formulating a hyperbolic pseudo-rotation matrix $H$ along the lines of Eq.~\ref{matrix}
\begin{equation}
\label{hmatrix}
H(\theta)=\begin{pmatrix}
\cosh \theta \hfill & -\sinh \theta \\
\sinh \theta \hfill & \cosh \theta
\end{pmatrix},
\end{equation}
to allow for hyperbolic rotation of any arbitrary vector.  Furthermore, our analysis of real rotation of a complex vector in Section~\ref{sec:rrotcv} can be applied easily here, as well.  Real hyperbolic pseudo-rotation of a complex vector amounts to hyperbolically rotating both the real and the imaginary components.

The trouble, however, with calling these transformations ``rotations'' is that none of the lengths $r$~(Eq.~\ref{rlength}), $h$~(Eq.~\ref{hlength}) or $s$~(Eq.~\ref{clength}) are conserved.  Furthermore, $\det H=\cosh 2\theta>1,\,\forall \theta\in\mathbb{R}$, showing the natural scaling properties of this mapping. Even if we change the sign on the top-right $\sinh$ of $H$ to ensure $\det H=1$, this merely satisfies a necessary, although insufficient condition for a rotation matrix.  
In any case, because hyperbolic pseudo-rotation fails the ``rotations preserve length'' test for \emph{all} lengths, it is not truly a rotation (hence ``pseudo-rotation'').  However, it still a very useful geometrical way to visualize the angle $\theta$ in the hyperbolic trig functions $\cosh\theta$, $\sinh\theta$ and the like, and will allow us to make sense of rotations through an imaginary angle.

\subsubsection{Hyperbolic rotation}
\begin{figure}
    \centering
		\includegraphics[width=0.9\textwidth]{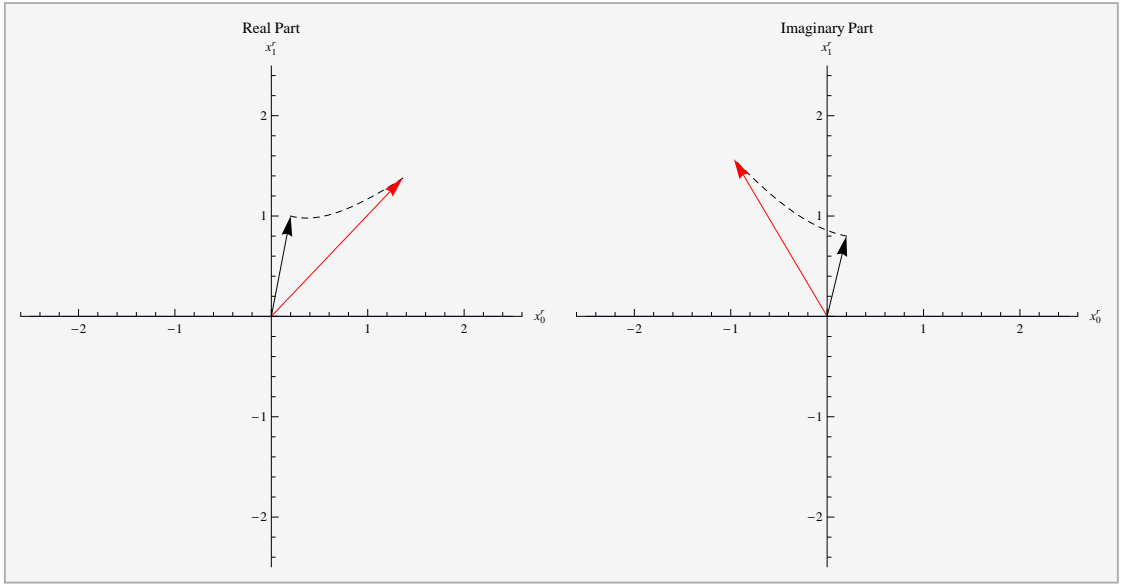}
    \caption{Visualizing imaginary rotation of a complex 2D vector $\vx=(4+2i,3+4i)$ through angle $\theta=i\pi/3$ by looking separately at the rotations of $\vec{x}\,^r=(4,3)$ in the $x_0^r$$x_1^r$-plane and $\vec{x}\,^i=(2,4)$ in the $x_0^i$$x_1^i$-plane.  Black arrow is original vector, red arrow is rotated vector, dotted line is path traced under rotation}
	\label{fig:crot}
\end{figure}

With all this out of the way, we can now finally piece together what actual hyperbolic rotations are.  Comparing the hyperbolic rotation matrix~(Eq.~\ref{cmatrix}) to the hyperbolic pseudo-rotation matrix~(Eq.~\ref{hmatrix}) we can see the only difference is the inclusion of the two `$i$'s on the $\sinh\theta$ terms.  If we carry through this mapping on a complex vector $\vx$ as given by Eq.~\ref{cvec}, we can simplify the resultant $\vx'$ (see~Appendix~\ref{dix:expand}):
\begin{subequations}
\begin{eqnarray}
\label{eq:xmap}
\vx'&=&R(i\theta)\begin{pmatrix}
x_0^r  \\
x_1^r \hfill
\end{pmatrix} \hfill + i\,R(i\theta)\begin{pmatrix}
x_0^i \\
x_1^i \hfill
\end{pmatrix} \\
\label{eq:hmap}
&=&M^{(r)} \begin{pmatrix}
\cosh\theta  \hfill \\
\sinh\theta  \hfill
\end{pmatrix} \hfill + i\,M^{(i)}\begin{pmatrix}
\cosh\theta  \hfill \\
\sinh\theta  \hfill
\end{pmatrix}
\end{eqnarray}
\end{subequations}
where
\begin{subequations}
\label{eq:ms}
\begin{eqnarray}
M^{(r)} &\equiv& \begin{pmatrix}
x_0^r \hfill & x_1^i \\
x_1^r \hfill & -x_0^i
\end{pmatrix}\\
M^{(i)} &\equiv& \begin{pmatrix}
x_0^i \hfill & -x_1^r \\
x_1^i \hfill & x_0^r
\end{pmatrix}
\end{eqnarray}
\end{subequations}
In other words, just as with real rotations, there are two, equally valid graphical ways of thinking about and visualizing this rotation through an imaginary angle.  The first~(Eq.~\ref{eq:xmap}) views it as hyperbolic rotation of the real and imaginary components of $\vx$ through an angle $\theta$ whereas the second~(Eq.~\ref{eq:hmap}) views it is a mapping of a portion of the unit hyperbola (up to angle $\theta$) due to the real and imaginary components of $\vx$.  The second one emphasizes that the path traced by the vector under this rotation is a hyperbola.  This is, of course, to be contrasted with the circle that a vector traces under real rotation. Graphically, we can see the equivalence of both both viewpoints, see Fig.~\ref{fig:crot}.

\begin{figure}
    \centering
		\includegraphics[width=0.9\textwidth]{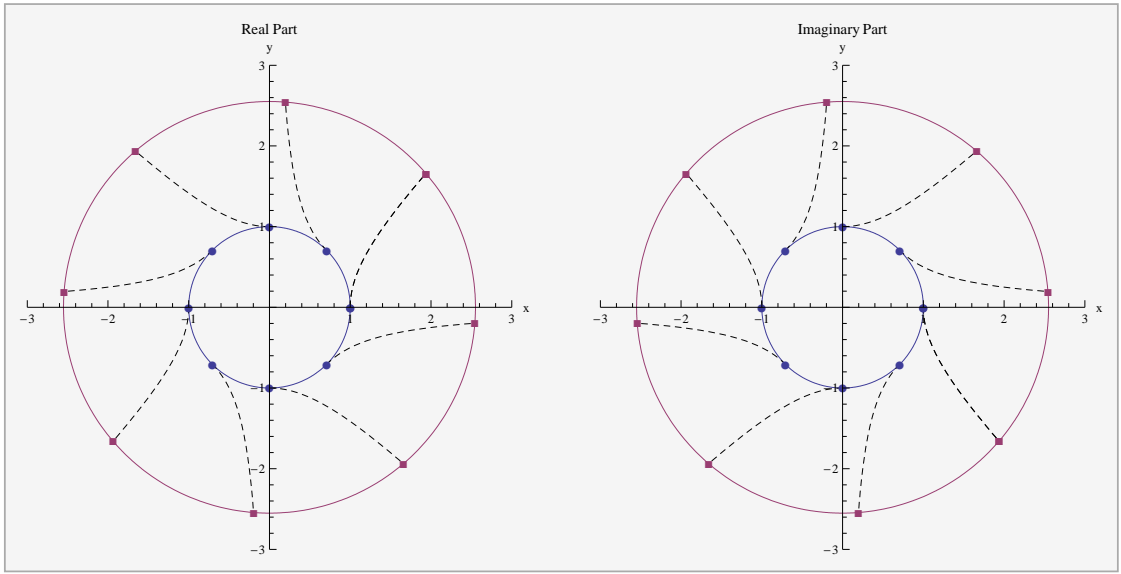}
    \caption{Imaginary rotation of the complex unit circle.  Dashed lines are paths of a few points along the circle traced under rotation}
	\label{fig:unit}
\end{figure}

Unlike with real rotation, however, the detailed reason for the conserved complex length, although easily shown analytically, is actually not so easy to intuit just from the picture. This is helped, somewhat, by looking at the transformation of the complex unit circle under imaginary rotation, see Fig.~\ref{fig:unit}.  Here, you can see a scaling happening, but both the real and complex parts of a vector scale the same way, but bend in opposite directions.  You can loosely argue that upon taking the inner product you essentially subtract these added parts from one another, and they cancel each other out.  Again, though, I have not fully worked out a way to see this directly from the graph.

I created a set of interactive tools to help with visualizing these mappings.  I've also included the other two visualization methods I described in Section~\ref{polar} for comparison.  These tools are available at: \href{http://tinyurl.com/imagrotate}{http://tinyurl.com/imagrotate}

\subsection{Orthogonal vs. Unitary Rotations}
\label{sec:unitary}
I've spent quite a bit of time discussing imaginary rotations which preserve the complex length $s$.  But the rotation matrix $R$~(Eq.~\ref{matrix}) is only one of a larger set (or group) of ``orthonormal matrices'' whose columns are orthogonal to one another and whose determinants are equal to one.  These other matrices include rotation matrices in higher dimensions, inversion and reflection matrices, and a few others.  There is a lot of formal mathematical discussion relating to these matrices and the relationship between them.\cite{arfken}  But the simple, \emph{graphical} relationship is that all these transformations preserve length and angle.

However, generally all discussion stops with discussion of real vectors and real lengths $r$.\cite{arfken}  What we've shown is that there's no problem defining a complex length which all of matrix members of the orthogonal group preserve when they operate on a complex vector.  Additionally, we've shown that these matrices can be complex as well - if we're careful. Thus we have a visual and graphical way of approaching the orthonormal group in a complex space.

There is, however, another set of matrices whose determinants are equal to one.  However, these so-called ``unitary matrices'' do not preserve the complex length $s$; instead they preserve the Hermitian length $h$~(Eq.~\ref{hlength}).  Thus, there is an important \emph{geometrical} difference between the matrices included in the orthogonal group and those in the unitary group, if we allow the orthogonal group to extend into complex space.  Additionally, we can use the graphical methods described earlier to help visualize the unitary operations as well, see Section~\ref{sec:pauli}.

\section{Higher Dimensions}
Although we've been dealing mostly with 2D vectors till this point, I would like to briefly discuss higher dimensions.  The rotation matrices take their standard form in higher dimensions.\cite{arfken} While the $Q$- and $M$-matrices (Eqs.~\ref{eq:qs} and~\ref{eq:ms}) will not appear in the same form in higher dimensions, the path traced by a vector by a rotation around an arbitrary axis will still be either a circle or a hyperbola in an hyperplane perpendicular to the axes of rotation.  Additionally, the visualizations developed are easily extended to three dimensions with two 3D plots, and I've included a visualization tool for this on the website: \href{http://tinyurl.com/imagrotate}{http://tinyurl.com/imagrotate}


\newcommand{\vrr}{{\boldsymbol{\vec{r}}}}
\section{Applications in Physics}
\subsection{Special Relativity}
\label{sec:relativity}
As I mentioned in the introduction, a natural application of this discussion is in the theory of special relativity.  A fundamental postulate of the theory is that the space-time interval
\begin{equation}
s^2=\vec{r}^{\,2}-\left(ct\right)^2
\end{equation}
between two events is invariant under a Lorentz transformation.\cite{goldstein3rd}  Here, $\vec{r}$ is an event's position in 3D space, $t$ is its measured time, and $c$ is the speed of light in vacuum.  It is natural to define a so-called ``4-vector'' 
\begin{equation}
\vrr=(i\,ct,x,y,z)
\end{equation}
such that
\begin{equation}
s^2=\ip{\vrr}
\end{equation}
is conserved.\cite{jackson1}  Well, this is just another way of saying that we want a transformation that preserves the complex length $s$ of the vector $\vrr$, and we know all about such transformations now!  These transformations are simply real and imaginary rotations.  For the sake of convenience, we'll just consider the 3D vector $\vx=(i\,ct,x,y)$ and see how it transforms.

Firstly, we can rotate $\vx$ about $i\,ct$ through a real angle.  Additionally, in principle, we could rotate $\vx$ about $x$ or $y$ through a real angle as well, but this isn't so exciting; we're needlessly messing up the simple distinction between time and position, and this really doesn't add any new physics.  But, let's see what happens when we rotate about $y$ through an imaginary angle $i\theta$.  Since $y$ will be unchanged under this rotation, we'll look only at the first two components and make use of Eqs.~\ref{eq:hmap} and \ref{eq:ms}.
\begin{eqnarray}
  \begin{pmatrix}
    i\,ct' \\
    x' \\
  \end{pmatrix}
&=&
\begin{pmatrix}
0 \hfill & 0 \\
x \hfill & -ct
\end{pmatrix} \begin{pmatrix}
\cosh\theta  \hfill \\
\sinh\theta  \hfill
\end{pmatrix} \hfill + i\begin{pmatrix}
ct \hfill & -x \\
0 \hfill & 0
\end{pmatrix}\begin{pmatrix}
\cosh\theta  \hfill \\
\sinh\theta  \hfill
\end{pmatrix}  \\
&=&
\begin{pmatrix}
    i\left(ct\cosh\theta-x\sinh\theta\right) \\
    x\cosh\theta-ct\sinh\theta \\
\end{pmatrix}\\
&=&
\begin{pmatrix}
    i\left(\gamma ct-\beta\gamma x\right) \\
    \gamma x-\beta\gamma ct\\
\end{pmatrix},
\end{eqnarray}
where $v/c\equiv\beta=\tanh\theta$, and $\gamma\equiv 1/\sqrt{1-\beta^2}$, where $v$ is the speed of one reference frame with respect to the other (along the x-axis).  Hence, $\gamma=\cosh\theta$, $\beta\gamma=\sinh\theta$. 
Generally this angle is called the ``rapidity.''  This result gives the standard Lorentz transformations:
\begin{subequations}
\begin{align}
ct'&=\gamma ct-\beta\gamma x\\
x'&=\gamma x - \beta\gamma ct
\end{align}
\end{subequations}
This approach towards this transformation allows us to think of the Lorentz transformation as an imaginary rotation of a specific form of complex vector where the angle of rotation depends on the velocity of one frame with respect to the other.  

The benefit of this analysis, then, is twofold. Firstly, for those of us who prefer to think in a complex Euclidean space, this approach offers an alternate viewpoint to the standard method of metrics, etc. On the other hand, for those who have no problem understanding Minkowski space, but have trouble visualizing complex spaces, this method offers a bridge into that mode of thought. Since these two approaches are mathematically equivalent, natural intuition in one frame should transform, with a little work, into intuition in the other.

\subsection{Visualizing Grassmann Numbers}
\label{sec:grassman}
Grassmann numbers (or variables), which arise in defining multiparticle propagators for fermions, are anti-commuting numbers.\cite{nairQFT} That is, for two numbers $a$ and $b$,
\begin{equation}
ab+ba=0.
\end{equation}
This weirdness is highlighted by setting $b=a$ for $a\neq 0$, giving
\begin{equation}
\label{eq:a20}
a^2=0,
\end{equation}
This latter property can be taken as a defining characteristic of these numbers, as well.  There are ways of realizing these quantities using matrices, however, using the methods developed in this paper, we have a vectorial way of realizing these numbers, with the added benefit that they can be visualized.

Just as 4-vectors in the Minkowski space can be visualized using a subset of the full space of complex vectors, we can visualize representations of Grassmann numbers using a different subset.  That is, if we find a vector $\vx$ such that its complex length $s=0$ we have found one of these Grassmann numbers.  Furthermore, we know that rotating this vector -- through real or imaginary angles -- preserves the length $s$, and thus we can find, and visualize, a whole range of these noncommuting numbers.

\begin{figure}
    \centering
		\includegraphics[width=0.9\textwidth]{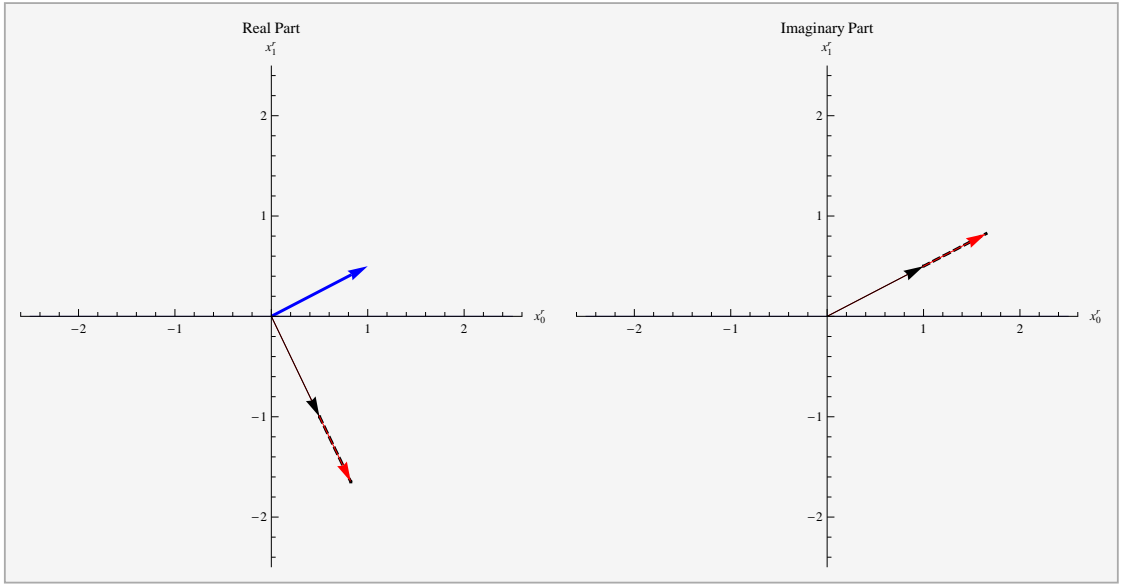}
    \caption{Visualizing Grassman numbers realized as complex vectors.  The thick blue arrow is the complex part of the vector as it would appear in the real plane.  For the imaginary rotation of a complex 2D Grassman vector $\vx=(1/2+i,-1+i/2)$ through angle $\theta=1/2$, the black arrow is original vector, red arrow is rotated vector, dotted line is path traced under rotation.}
	\label{fig:grass}
\end{figure}

Any complex vector of the form
\begin{equation}
\vx=\begin{pmatrix}
\pm \alpha \mp i\,\beta \\
\beta+i\,\alpha
\end{pmatrix}
\end{equation}
has $s=0$, for $\alpha$ and $\beta$ real.  This vector has the additional property that the real lengths of the real and imaginary parts of $\vx$ are equivalent.  Note, then, that the Hermitian length of $\vx$ is two times the length of either the real or imaginary components of the vector.  This follows the traditional notion of commuting numbers, i.e. $ab+ab=2ab$.

Additionally, if we consider the graphical interpretation of an inner product as a ``measure of parallelity,'' then Eq.~\ref{eq:a20} says that ``$\vx$ is perpendicular to itself.''  Well, looking at Fig.~\ref{fig:grass}, graphically $\vx$ is, in a way, perpendicular to itself.  If you put the real and complex parts of the vector on the same plane they would be orthogonal.  


Furthermore, the effects of the rotation matrix for real angles is the same as for any vector.  However, for imaginary angles $i\theta$, it simply scales the vector uniformly by a factor of $(\cosh\theta+\sinh\theta)$, see Fig.~\ref{fig:grass} (easily seen algebraically using Eq.~\ref{eq:hmap}).  
Thus we can go from any one of these vectors to another by a real and imaginary rotation.  So we can represent Grassmann numbers by any complex vector whose real and imaginary parts are the same length and are orthogonal to each other in real space.

\subsection{Pauli Spin Matrices}
\label{sec:pauli}
In quantum mechanics we are generally not interested orthogonal transformations which preserve $s$, but unitary transformations which preserve $h$.  Although for the majority of this paper I have been analyzing the former class of transformations, the visualization tools developed can be used for the latter as well.  Furthermore, the clear graphical distinction that arises highlights the physical differences between unitary rotations and orthogonal rotations.

As an example, we'll consider particles with spin-$\frac{1}{2}$ which can be represented by a complex 2D spinor vector $\vec{\chi}=(\Psi_{up},\Psi_{down})$.  Here, $\Psi$ represents the complex wavefunction of the particle which is itself a function of space.  Upon rotating the coordinate system, $\Psi_{up}$ gets blended into $\Psi_{down}$ in manner preserving the Hermetian length $h$ of $\vec{\chi}$.  The precession of $\vec{\chi}$ under rotations of space is mapped using the three Pauli spin matrices:
\begin{subequations}
\begin{align}
\sigma_x&=
\begin{pmatrix}
0 \hfill & 1 \\
1 \hfill & 0
\end{pmatrix}\\
\sigma_y&=
\begin{pmatrix}
0 \hfill & -i \\
i \hfill & 0
\end{pmatrix}\\
\sigma_z&=
\begin{pmatrix}
1 \hfill & 0 \\
0 \hfill & -1
\end{pmatrix}
\end{align}
\end{subequations}
which are called the ``generators'' of the rotations.  That is, for a rotation of space through angle $\theta$ around the axis pointing along the unit-vector $\hat{n}=\hat{\imath}n_x+\hat{\jmath}n_y+\hat{k}n_z$, the general precession operator $\mathcal{R}$ is given by:\cite{schwabl}
\begin{equation}
\mathcal{R}=\exp\left[i\theta (\vec{\sigma}\cdot\hat{n})/2\right],
\end{equation}
where $\vec{\sigma}=\hat{\imath}\sigma_x+\hat{\jmath}\sigma_y+\hat{k}\sigma_z$, so that
\begin{equation}
\vec{\chi}\,'=\mathcal{R}(\hat{n},\theta)\,\vec{\chi}
\end{equation}

There are a few tricky things to deal with here.  First, even though we are describing the effect of rotations through 3D space, the dimensionality of the spin matrices requires that they act on a two-dimensional vector ($\vec{\chi}$) which is a function of space.
Secondly, although physics texts generally analyze the algebraic (commutation) properties of the $\sigma$ matrices, in most of the ``standard texts'' the actual form of the resultant operator does not appear.

This is strange, because this operator is actually quite simple:
\begin{equation}
\label{eq:rfull}
\mathcal{R}=
\begin{pmatrix}
\cos\frac{\theta}{2} +i n_z \sin\frac{\theta}{2} \hfill & i ( n_x - i n_y) \sin\frac{\theta}{2} \\
i ( n_x + i n_y) \sin\frac{\theta}{2} \hfill & \cos\frac{\theta}{2} - i n_z \sin\frac{\theta}{2}
\end{pmatrix}.
\end{equation}
There is one thing we can quickly note about this matrix.  For rotations of space through angle $\theta$ around $-\hat{\jmath}$ this matrix reduces to the standard rotation matrix Eq.~\ref{matrix}, $\mathcal{R}(\theta)=R(\frac{\theta}{2})$.  Thus, the spinor traces a circle at half the rate of ordinary rotation, and to get back to the initial state, we need to rotate space around $-\hat{\jmath}$ through an angle $2\times 2\pi$.

Furthermore, although this is a unitary transformation, the visualization tools developed in this paper can be used to graph the precession of the complex spinor due to various spacial rotations.  I've also included a visualization for this on the website, as well:\\ \href{http://tinyurl.com/imagrotate}{http://tinyurl.com/imagrotate}

%
%
%
%

\section{Topics for future analysis}
I've only really discussed easy rotations around a vector perpendicular to the $x_0x_1$-plane.  It would be interesting to generalize all 3D rotations via complex Euler angles and see how those work.  Additionally, allowing a rotation though a more general complex angle $\Theta=\phi+i\,\theta$ should account for all rotations, and it would also be nice to see how those mappings look graphically.  Furthermore, the graphical interpretation of an inner product as a ``measure of parallelity'' can be looked into further with complex vectors.  The relation between angle and the complex inner product should also be analyzed, and compared to the various forms of angles in complex vector spaces.\cite{complexspaceangle}  Additionally, the polar representation that I alluded to in Section~\ref{polar} can also be expounded upon graphically, with its applications.
\newpage

\appendix
\section{Expansion of hyperbolic rotation}
\label{dix:expand}
We expand:
\begin{eqnarray}
\vx'&=&R(i\theta)\vx\\
&=&\begin{pmatrix}
\cosh \theta \hfill & -i \sinh \theta \\
i \sinh \theta \hfill & \cosh \theta
\end{pmatrix}\begin{pmatrix}
x_0^r  \\
x_1^r \hfill
\end{pmatrix}+i\, \begin{pmatrix}
\cosh \theta \hfill & -i \sinh \theta \\
i \sinh \theta \hfill & \cosh \theta
\end{pmatrix}\begin{pmatrix}
x_0^i \\
x_1^i \hfill
\end{pmatrix}\\
&=&\begin{pmatrix}
x_0^r\cosh \theta   -i x_1^r\sinh \theta \\
i x_0^r\sinh \theta  + x_1^r\cosh \theta
\end{pmatrix}+i\, \begin{pmatrix}
x_0^i\cosh \theta   -i x_1^i\sinh \theta \\
i x_0^i\sinh \theta  + x_1^i\cosh \theta
\end{pmatrix}\\
&=&\begin{pmatrix}
x_0^r\cosh \theta   -i x_1^r\sinh \theta \\
i x_0^r\sinh \theta  + x_1^r\cosh \theta
\end{pmatrix}+\begin{pmatrix}
i\, x_0^i\cosh \theta   + x_1^i\sinh \theta \\
-x_0^i\sinh \theta  +i\,x_1^i\cosh \theta
\end{pmatrix}\\
&=&
\label{fullxp}
\begin{pmatrix}
x_0^r\cosh \theta   -i x_1^r\sinh \theta +i\, x_0^i\cosh \theta   + x_1^i\sinh \theta\\
i x_0^r\sinh \theta  + x_1^r\cosh \theta-x_0^i\sinh \theta  +i\,x_1^i\cosh \theta
\end{pmatrix}\\
&=&\begin{pmatrix}
x_0^r\cosh \theta   + x_1^i\sinh \theta\\
x_1^r\cosh \theta-x_0^i\sinh \theta \\
\end{pmatrix}+i\,
\begin{pmatrix}
x_0^i\cosh \theta-x_1^r\sinh \theta  \\
x_1^i\cosh \theta+x_0^r\sinh \theta
\end{pmatrix}
\\
&=&\begin{pmatrix}
x_0^r \hfill & x_1^i \\
x_1^r \hfill & -x_0^i
\end{pmatrix}
\begin{pmatrix}
\cosh \theta \\
\sinh \theta \\
\end{pmatrix}+i\,\begin{pmatrix}
x_0^i \hfill & -x_1^r \\
x_1^i \hfill & x_0^r
\end{pmatrix}
\begin{pmatrix}
\cosh \theta \\
\sinh \theta \\
\end{pmatrix}
\end{eqnarray}

\section{Proof of conserved complex length with modified hyperbolic rotations}
\label{dix:cons}
Consider a $n\times n$ matrix with matrix $R(i\theta)$ (Eq.~\ref{cmatrix}) in the top-left corner, $1$'s along the rest of the $n-2$ diagonals, and $0$ elsewhere.  This corresponds to a rotation in the $x_0x_1$-plane.  In principle, we could have chosen an arbitrary position along the diagonal to place $R(i\theta)$, or a similar form of the rotation matrix two rotate in any arbitrary hyperplane.  Since in all these cases the other diagonals are equal to one, there is no change of the $n-2$ lengths, and we can then concentrate on the two rotating dimensions.  If complex length is conserved in these dimensions, the total complex length is then also conserved.

So we'll use Eq.~\ref{fullxp}:
\begin{align}
\ip{\vx'}={}&  \left(x_0^r\cosh \theta   -i x_1^r\sinh \theta +i\, x_0^i\cosh \theta  +x_1^i\sinh \theta\right)^2 \\
\notag&+\left(i x_0^r\sinh \theta  + x_1^r\cosh \theta-x_0^i\sinh \theta  +i\,x_1^i\cosh \theta\right)^2\\
={}&[-(x_0^i)^2 \cosh^2\theta + 2 i\, x_0^i x_0^r \cosh^2\theta +
 (x_0^r)^2 \cosh^2\theta +\\
\notag &+
 2 i\, x_0^i x_1^i \cosh\theta \sinh\theta +
 2 x_1^r x_1^i \cosh\theta \sinh\theta +\\
\notag &+
 2 x_0^i x_1^r \cosh\theta \sinh\theta -
 2 i\, x_1^r x_1^r \cosh\theta \sinh\theta +\\
\notag &+
 (x_1^i)^2 \sinh^2\theta - 2 i\, x_1^i x_1^r \sinh^2\theta -
 (x_1^r)^2 \sinh^2\theta]\\
\notag &+\\
\notag &[- (x_1^i)^2  \cosh^2\theta + 2 i\,x_1^i x_1^r \cosh^2\theta +
 (x_1^r)^2 \cosh^2\theta +\\
\notag &- 2 i\,x_0^i x_1^i \cosh\theta \sinh\theta -
 2 x_0^r x_1^i \cosh\theta \sinh\theta +\\
\notag &- 2 x_0^i x_1^r \cosh\theta \sinh\theta +
 2 i\,x_0^r x_1^r \cosh\theta \sinh\theta +\\
\notag &+
 (x_0^i)^2 \sinh^2\theta - 2 i\,x_0^i x_0^r \sinh^2\theta -
 (x_0^r)^2 \sinh^2\theta]\\
={}&
 [-(x_0^i)^2 \cosh^2\theta + 2 i\, x_0^i x_0^r \cosh^2\theta +
 (x_0^r)^2 \cosh^2\theta +\\
\notag &+
 (x_1^i)^2 \sinh^2\theta - 2 i\, x_1^i x_1^r \sinh^2\theta -
 (x_1^r)^2 \sinh^2\theta]\\
\notag &+\\
\notag &[- (x_1^i)^2  \cosh^2\theta + 2 i\,x_1^i x_1^r \cosh^2\theta +
 (x_1^r)^2 \cosh^2\theta +\\
\notag &+
 (x_0^i)^2 \sinh^2\theta - 2 i\,x_0^i x_0^r \sinh^2\theta -
 (x_0^r)^2 \sinh^2\theta]\\
={}&
 (x_0^r)^2 (\cosh^2\theta  -
 \sinh^2\theta)-(x_0^i)^2(\cosh^2\theta- \sinh^2\theta)
+\\
\notag &+(x_1^r)^2 (\cosh^2\theta-\sinh^2\theta)-(x_1^i)^2 (\cosh^2\theta-\sinh^2\theta)+ \\
\notag &+ 2 i\, x_0^i x_0^r (\cosh^2\theta  - \sinh^2\theta)
 + 2 i\,x_1^i x_1^r (\cosh^2\theta  - \sinh^2\theta)\\
 ={}&
 (x_0^r)^2 + 2 i\, x_0^i x_0^r -(x_0^i)^2+(x_1^r)^2 + 2 i\,x_1^i x_1^r -(x_1^i)^2 \\
  ={}&
 (x_0^r)^2 + 2 i\, x_0^i x_0^r +(i\,x_0^i)^2+(x_1^r)^2 + 2 i\,x_1^i x_1^r +(i\, x_1^i)^2 \\
  ={}&
 (x_0^r+ i\, x_0^i )^2+(x_1^r + i\,x_1^i)^2 \\
 ={}&\ip{\vx}
\end{align}


\begin{acknowledgments}
I want to thank Avi Ziskind for his very helpful, and insightful suggestions through all stages of writing this paper.
\end{acknowledgments}

\bibliography{blog}

\begin{thebibliography}{10}

\bibitem{jackson1}
J.D. Jackson.
\newblock {\em Classical Electrodynamics}.
\newblock John Wiley \& Sons, Inc., 1st edition, 1966.

\bibitem{lay}
D.C. Lay.
\newblock {\em Linear Algebra and Its Applications}.
\newblock Addison-Wesley, Reading, MA, 3rd edition, 2003.

\bibitem{gibbs2}
J.~Willard Gibbs.
\newblock {\em The scientific papers of J. Willard Gibbs}, volume~II.
\newblock Longmans, Green, and Co., London, England, 1906.

\bibitem{larson}
R.~Larson and B.H. Edwards.
\newblock {\em Elementary Linear Algebra}.
\newblock Houghton Mifflin Company, 4th, online chapters edition, 1999.

\bibitem{tensorgeom}
C.T.J. Dodson and T.~Poston.
\newblock {\em Tensor Geometry: The Geometric Viewpoint and its Uses}.
\newblock Springer, 2nd edition, 1997.

\bibitem{needham}
Tristan Needham.
\newblock {\em Visual Complex Analysis}.
\newblock Oxford University Press, Oxford, UK, 1997.

\bibitem{arfken}
G.B. Arfken and H.J. Weber.
\newblock {\em Mathematical Methods for Physicists}.
\newblock Elsevier, 6th edition, 2005.

\bibitem{goldstein3rd}
H.~Goldstein, C.~Poole, and J.~Safko.
\newblock {\em Classical Mechanics}.
\newblock Cambridge University Press, San Francisco, CA, 3rd edition, 2002.

\bibitem{nairQFT}
V.~P. Nair.
\newblock {\em Quantum field theory: A modern perspective}.
\newblock Springer, USA, 1st edition, 2005.

\bibitem{schwabl}
F.~Schwabl.
\newblock {\em Quantum Mechanics}.
\newblock Springer, 3rd edition, 2005.

\bibitem{complexspaceangle}
K.~Scharnhorst.
\newblock Angles in complex vector spaces.
\newblock {\em Acta Applicandae Mathematicae: An International Survey Journal
  on Applying Mathematics and Mathematical Applications}, 69(1):95--103,
  October 2001.

\end{thebibliography}
\bibliographystyle{unsrturl}

%
%

\end{document}